\documentclass[11pt]{article}
\usepackage{latexsym,amsmath,amsfonts}
\begin{document}
\begin{center}
{\LARGE\textbf{Finite-time consensus for nonlinear multi-agent systems with fixed topologies}}\\
\bigskip
\bigskip
Yilun Shang\footnote{Department of Mathematics, Shanghai Jiao Tong University, Shanghai 200240, CHINA. email: \texttt{shyl@sjtu.edu.cn}}\\
\end{center}

\begin{abstract}
In this paper, we study finite-time state consensus problems for
continuous nonlinear multi-agent systems. Building on the theory of
finite-time Lyapunov stability, we propose sufficient criteria which
guarantee the system to reach a consensus in finite time, provided
that the underlying directed network contains a spanning tree. Novel
finite-time consensus protocols are introduced as examples for
applying the criteria. Simulations are also presented to illustrate
our theoretical results.

\bigskip

\end{abstract}

\bigskip
\normalsize

\noindent{\Large\textbf{1. Introduction}}
\smallskip

Distributed coordination for multi-agent systems has become an
active research topic and attracted great attention of researchers
in recent years; see e.g. \cite{13,1,6,2,3,5,45}. A typical problem
in the area is agreement or consensus problem, which means to design
a network protocol based on the local information obtained by each
agent, such that all agents finally reach an agreement on certain
quantities of interest. The network protocol is an interaction rule,
which ensures the whole group can achieve consensus on the shared
data in a distributed manner. Consensus problems cover a very broad
spectrum of applications including formation control, distributed
filtering, multi-sensor data fusion, and distributed computation, to
cite but a few examples. We refer the reader to the survey papers
\cite{4,12} and references therein.

In the study of consensus problems, convergence rate is an important
index to evaluate the proposed protocol. Most of the existing
protocols (including those appeared in the aforementioned works) can
not result in state consensus in a finite time, that is, consensus
is only achieved asymptotically. Hence, finite-time consensus is
more appealing and there are a number of settings where finite-time
convergence is a desirable property. Recently, finite-time consensus
problems have attracted the attention of some researchers. Some
related works are briefly reviewed as follows. \cite{10} introduces
the signed gradient descent flows which serve as discontinuous
protocols for finite-time coordination under connected undirected
topologies. Two discontinuous distributed algorithms are
characterized in \cite{17} to achieve, respectively, max and min
consensus in finite time over strongly connected digraphs. Several
classes of continuous nonlinear protocols coming out of the typical
linear protocol in \cite{2} are considered by \cite{8,7}. The
authors show that they are efficient finite-time agreement protocols
provided the directed interconnection topology has a spanning tree.
The question of having communication delays is further discussed in
\cite{40}. \cite{16,9} deal with finite-time consensus under a
general framework for finite-time semistability of homogeneous
systems, and the underlying topology is assumed to be a connected
undirected graph. A continuous finite-time tracking control problem
is investigated in \cite{41} for a non-holonomic wheeled mobile
robot by carefully selecting control gains.

In this note, we aim to identify general criteria for solving
finite-time consensus problems with continuous protocols under
directed weighted fixed topologies. Based on the theory of
finite-time Lyapunov stability \cite{14,15}, we show that under
protocols satisfying our conditions, the states of agents reach a
consensus in finite time when the interaction topology has a
directed spanning tree. Novel protocols are proposed as a proof of
the criteria, and we corroborate their finite-time convergence
property thereby.

The rest of the paper is organized as follows. In Section 2, we
provide some preliminaries and formulate the finite-time convergence
criteria. Section 3 contains the convergence analysis under directed
fixed topologies. Some concrete examples with numerical simulations
are given in Section 4 and we draw conclusion in Section 5.

\bigskip
\noindent{\Large\textbf{2. Problem formulation}}
\smallskip

In general, information exchange between agents in a multi-agent
system can be modeled by directed graphs \cite{11,2}. Before we
proceed, we first introduce some basic concepts and notions in graph
theory.

Let
$\mathcal{G}(\mathcal{A})=(\mathcal{V}(\mathcal{G}),\mathcal{E}(\mathcal{G}),\mathcal{A})$
be a weighted directed graph with the set of vertices
$\mathcal{V}(\mathcal{G})=\{1,2,\cdots,n\}$ and the set of arcs
$\mathcal{E}(\mathcal{G})\subseteq\mathcal{V}(\mathcal{G})\times\mathcal{V}(\mathcal{G})$.
The vertex $i$ in $\mathcal{G}(\mathcal{A})$ represents the $i$th
agent, and a directed edge $(i,j)\in\mathcal{E}(\mathcal{G})$ means
that agent $j$ can directly receive information from agent $i$, the
parent vertex. The set of neighbors of vertex $i$ is denoted by
$\mathcal{N}(\mathcal{G},i)=\{j\in\mathcal{V}(\mathcal{G})|\
(j,i)\in\mathcal{E}(\mathcal{G})\}$.
$\mathcal{A}=(a_{ij})\in\mathbb{R}^{n\times n}$ is called the
weighted adjacency matrix of $\mathcal{G}(\mathcal{A})$ with
nonnegative elements and $a_{ij}>0$ if and only if
$j\in\mathcal{N}(\mathcal{G},i)$. The corresponding graph Laplacian
$L(\mathcal{A})=(l_{ij})\in\mathbb{R}^{n\times n}$ can be defined as
$$
l_{ij}=\left\{\begin{array}{ll}\sum_{k=1,k\not=n}^{n}a_{ik},&j=i\\
-a_{ij},&j\not=i
\end{array}\right..
$$
If $\mathcal{A}^T=\mathcal{A}$, we say $\mathcal{G}(\mathcal{A})$ is
undirected. As is known, the Laplacian matrix of undirected graph is
positive semidefinite.

A directed tree is a directed graph with one root vertex which has
no parent vertex, every other vertex has exactly one parent, and the
root can be connected to any other vertices through directed paths.
A spanning tree of a directed graph $\mathcal{G}$ is a directed tree
which is a spanning subgraph. A directed graph $\mathcal{G}$ is
called strongly connected if there is a directed path from $i$ to
$j$ between any two distinct vertices
$i,j\in\mathcal{V}(\mathcal{G})$. An undirected graph is connected
if it is strongly connected when regarded as a directed graph. A
strongly connected component of a directed graph is an induced
subgraph that is maximal, subject to being strongly connected. As is
known, the strongly connected components of a given directed graph
partition its vertex set.

Here, we consider a system consisting of $n$ autonomous agents,
indexed by $1,2,\cdots,n$. The information interaction topology
among them are described by the weighted directed graph
$\mathcal{G}(\mathcal{A})$ as defined above. We further assume the
diagonal entries of $\mathcal{A}$ are zeroes. The continuous-time
dynamics of $n$ agents is described as follows:
\begin{equation}
\dot{x}_i(t)=u_i(t),\quad i=1,2,\cdots,n,\label{1}
\end{equation}
where $x_i(t)\in\mathbb{R}$ is the state of the $i$th agent, and
$u_i(t)\in\mathbb{R}$ is the state feedback, called protocol, to be
designed. Denote $x(t)=(x_1(t),\cdots,x_n(t))^T$ and
$1=(1,\cdots,1)^T$ with compatible dimensions. For a vector
$z\in\mathbb{R}^n$, let $\|z\|_{\infty}$ denote its
$l^{\infty}$-norm, for a matrix $Z\in\mathbb{R}^{n\times n}$, let
$\|Z\|_{\infty}$ denote its induced $l^{\infty}$-norm, and for a
number $z\in\mathbb{R}$, let $|z|$ denote its absolute value.

Given protocol $\{u_i : i=1,2,\cdots,n\}$, the multi-agent system is
said to solve a consensus problem if for any initial states and any
$i,j\in\{1,\cdots,n\}$, $|x_i(t)-x_j(t)|\rightarrow0$ as
$t\rightarrow\infty$ (c.f. \cite{2}); and it is said to solve a
finite-time consensus problem if for any initial states, there is
some finite-time $t^*$ such that $x_i(t)=x_j(t)$ for any
$i,j\in\{1,\cdots,n\}$ and $t\ge t^*$ (c.f. \cite{8}).

We now present our protocol as follows:
\begin{equation}
u_i=f_i\bigg(\sum_{j\in\mathcal{N}(\mathcal{G}(\mathcal{A}),i)}a_{ij}(x_j-x_i)\bigg),\label{2}
\end{equation}
where functions $f_i:\mathbb{R}\rightarrow\mathbb{R}$,
$i=1,\cdots,n$, satisfy the following two assumptions, which will be
shown as sufficient criteria for finite-time consensus:\\
\textbf{(A1)}\quad For $i=1,\cdots,n$, $f_i$ is a continuous and
increasing function
with $f_i(z)=0$ if and only if $z=0$.\\
\textbf{(A2)}\quad Given the interaction topology
$\mathcal{G}(\mathcal{A})$ and initial state $x(0)$, there exist
some constants $\beta>0$ and $0<\alpha<1$ such that, for any
$0<|z|\le\|L(\mathcal{A})\|_{\infty}\|x(0)\|_{\infty}$,
\begin{equation}
\min_{1\le i\le
n}\frac{f_i(z)^2}{\big(\int_0^zf_i(s)\mathrm{d}s\big)^{\alpha}}\ge\beta.\label{3}
\end{equation}

We give two remarks here.

\noindent\textbf{Remark 1.}\itshape \quad The continuity in
Assumption (A1) is meant to guarantee the existence of solutions of
differential equations (\ref{1}) on $[0,\infty)$ for any initial
value $x(0)$, as is indicated by Peano's Theorem (e.g. \cite{18}
pp.10). \normalfont

\noindent\textbf{Remark 2.}\itshape \quad It is easy to see that the
linear protocol proposed in \cite{2} (i.e. by setting $f_i(x)=kx$
for $k>0$) does not satisfy Assumption (A2). In fact, consensus can
never occur in a finite time for such linear protocols.\normalfont

\bigskip
\noindent{\Large\textbf{3. Convergence analysis}}
\smallskip

In this section, the convergence property of the consensus protocol
(\ref{2}) for multi-agent system (\ref{1}) is given. Prior to the
establishment, we introduce the following two lemmas regarding the
Laplacian matrix $L(\mathcal{A})$.

\smallskip
\noindent\textbf{Lemma 1.}\ \cite{2,3,8}\itshape \quad Assume
$\mathcal{G}(\mathcal{A})$ is a
directed graph with Laplacian matrix $L(\mathcal{A})$, then we have\\
(i)\quad $L(\mathcal{A})1=0$ and all non-zero eigenvalues have
positive real
parts;\\
(ii)\quad $L(\mathcal{A})$ has exactly one zero eigenvalue if and
only if $\mathcal{G}(\mathcal{A})$ has a spanning tree;\\
(iii)\quad If $\mathcal{G}(\mathcal{A})$ is strongly connected, then
there is a positive column vector $\omega\in\mathbb{R}^n$ such that
$\omega^TL(\mathcal{A})=0$;\\
(iv)\quad Let $b=(b_1,\cdots,b_n)^T$ be a nonnegative vector and
$b\not=0$. If $\mathcal{G}(\mathcal{A})$ is undirected and
connected, then $L(\mathcal{A})+diag(b)$ is positive definite. Here,
$diag(b)$ is the diagonal matrix with the $(i,i)$ entry being
$b_i$.\normalfont
\smallskip

\noindent\textbf{Lemma 2.}\ \cite{8}\itshape \quad Suppose
$\mathcal{G}(\mathcal{A})$ is strongly connected, and $\omega$ is
given as in Lemma 1. Then
$\mathrm{diag}(\omega)L(\mathcal{A})+L(\mathcal{A})^T\mathrm{diag}(\omega)$
is the graph Laplacian of the connected undirected graph
$\mathcal{G}(\mathrm{diag}(\omega)\mathcal{A}+\mathcal{A}^T\mathrm{diag}(\omega))$.
\normalfont\smallskip

In what follows we present our main result.

\smallskip
\noindent\textbf{Theorem 1.}\itshape \quad If the interaction
topology $\mathcal{G}(\mathcal{A})$ has a spanning tree, then the
system (\ref{1}) solves a finite-time consensus problem when
protocol (\ref{2}) is applied.\normalfont

\smallskip
\noindent\textbf{Proof}. We prove the theorem through the following
three steps.

Step 1. Suppose that $\mathcal{G}(\mathcal{A})$ is strongly
connected.

By Lemma 1, there exists a positive vector
$\omega=(\omega_1,\cdots,\omega_n)^T\in\mathbb{R}^n$ such that
$\omega^TL(\mathcal{A})=0$. Let $y_i=\sum_{j=1}^na_{ij}(x_j-x_i)$
and $y=(y_1,\cdots,y_n)^T$. Therefore, $y=-L(\mathcal{A})x$, $y\bot\
\omega$ and $\dot{x}_i=f_i(y_i)$. Let $f=(f_1,\cdots,f_n)^T$, and
then we may rewrite the system in a compact form as $\dot{x}=f(y)$.
We define a Lyapunov function as:
$$
V(t)=\sum_{i=1}^n\omega_i\int_0^{y_i}f_i(s)\mathrm{d}s.
$$
Obviously, $V(t)\ge0$, and $V(t)=0$ if and only if $y(t)=0$.
Differentiating $V(t)$, we get
$$
\frac{\mathrm{d}V(t)}{\mathrm{d}t}=\sum_{i=1}^n\omega_if_i(y_i)\dot{y}_i
=-f(y)^T\mathrm{diag}(\omega)L(\mathcal{A})f(y).
$$

Denote
$B=\big(\mathrm{diag}(\omega)L(\mathcal{A})+L(\mathcal{A})^T\mathrm{diag}(\omega)\big)\big/2$.
By Lemma 2, $B$ can be regarded as a Laplacian matrix of an
connected undirected graph and hence is positive semidefinite.
Suppose $V(t)\not=0$, namely, $y\not=0$. We obtain
\begin{eqnarray}
\frac{\mathrm{d}V(t)}{\mathrm{d}t}&=&-\frac12f(y)^T\big(\mathrm{diag}(\omega)L(\mathcal{A})+L(\mathcal{A})^T\mathrm{diag}(\omega)\big)f(y)\nonumber\\
&=&-\frac{f(y)^TBf(y)}{f(y)^Tf(y)}\cdot\frac{f(y)^Tf(y)}{V(t)^{\alpha}}\cdot
V(t)^{\alpha},\label{4}
\end{eqnarray}
where $\alpha\in(0,1)$ is defined in Assumption (A2).

Consider the first quality in the right-hand side of equality
(\ref{4}). Let $\mathcal{S}=\{\xi\in\mathbb{R}^n:\xi^T\xi=1 \
\mathrm{and\ the\ nonzero\ terms\ of\ } \xi_1,\cdots,\xi_n\
\mathrm{are\ not\ with\ the\ same\ sign}\}$. Then $\mathcal{S}$ is a
bounded closed set. Since $\xi^TB\xi$ is a continuous function and
for any $\xi\in\mathcal{S}$, $\xi^TB\xi>0$ (involving Lemma 1 and
the positive semidefiniteness of $B$), we have that
$\min_{\xi\in\mathcal{S}}\xi^TB\xi:=C_1>0$. Thereby
$$
\frac{f(y)^TBf(y)}{f(y)^Tf(y)}=\frac{f(y)^T}{\sqrt{f(y)^Tf(y)}}B\frac{f(y)}{\sqrt{f(y)^Tf(y)}}\ge
C_1.
$$

Note that
$$
0<|y_i(t)|\le\|y(t)\|_{\infty}=\|-L(\mathcal{A})x(t)\|_{\infty}\le
\|L(\mathcal{A})\|_{\infty}\|x(t)\|_{\infty}\le
\|L(\mathcal{A})\|_{\infty}\|x(0)\|_{\infty},
$$
where the last inequality follows from the fact that
$\|x(t)\|_{\infty}$ is non-increasing. By exploiting
$C_r$-inequality and (\ref{3}) in Assumption (A2), we get
$$
\frac{f(y)^Tf(y)}{V(t)^{\alpha}}=\frac{\sum_{i=1}^nf_i(y_i)^2}{\big(\sum_{i=1}^n\omega_i\int_0^{y_i}f_i(s)\mathrm{d}s\big)^{\alpha}}\ge
\frac{\sum_{i=1}^nf_i(y_i)^2}{\sum_{i=1}^n\omega_i^{\alpha}\big(\int_0^{y_i}f_i(s)\mathrm{d}s\big)^{\alpha}}\ge
C_2\beta,
$$
where $C_2=1/\max_{1\le i\le n}\omega_i^{\alpha}>0$. Combing these
with Equation (\ref{4}) yields
$$
\frac{\mathrm{d}V(t)}{\mathrm{d}t}\le-C_1C_2\beta V(t)^{\alpha}.
$$
Consider the differential equation
$$
\frac{\mathrm{d}v(t)}{\mathrm{d}t}=-C_1C_2\beta v(t)^{\alpha}
$$
with initial value $v(0)=V(0)$, and its unique solution is shown to
be given by
$$
v(t)=\left\{\begin{array}{lc}\big(-C_1C_2\beta(1-\alpha)t+V(0)^{1-\alpha}\big)^{\frac1{1-\alpha}},&t<t^*\\
0,&t\ge t^*
\end{array}\right.
$$
where $t^*=V(0)^{1-\alpha}\big/C_1C_2\beta(1-\alpha)$. By Comparison
Principle of differential equations (e.g. \cite{18} pp.26), we have
$V(t)\le v(t)$. Consequently, $V(t)$ and $y(t)$ approach zero in
finite time $t^*$. Since $y=-L(\mathcal{A})x$, $y=0$ implies that
$x\in\mathrm{span}\{1\}=\{c1:c\in\mathbb{R}\}$ and $\dot{x}(t)=0$ by
using Lemma 1 and Assumption (A1). Hence the system solves a
finite-time consensus problem.

Step 2. Suppose that $\mathcal{G}(\mathcal{A})$ has a spanning tree
with root vertex $i$, and the subgraph induced by the remaining
vertices is strongly connected. Moreover, we suppose there exists no
directed path connecting those vertices to $i$.

Without loss of generality, assume that the root vertex $i$ is
vertex $n$. From the protocol (\ref{2}), we see the state $x_n$ is
time-invariant, and $a_{n1}=\cdots=a_{nn}=0$. Let $b_i=a_{in}$ for
$1\le i\le n-1$, $\widetilde{b}=(b_1,\cdots,b_{n-1})^T$ and
$\widetilde{\mathcal{A}}=(a_{ij})_{1\le i,j\le
n-1}\in\mathbb{R}^{(n-1)\times (n-1)}$. By our assumption,
$\widetilde{b}\not=0$. Denote $z_i=x_i-x_n$ for $i=1,\cdots,n-1$,
and $z=(z_1,\cdots,z_{n-1})^T$. Then for $i=1,\cdots,n-1$,
$$
\dot{z}_i=\dot{x}_i=f_i\bigg(\sum_{j=1}^{n-1}a_{ij}(z_j-z_i)-b_iz_i\bigg).
$$
Let $y_i=\sum_{j=1}^{n-1}a_{ij}(z_j-z_i)-b_iz_i$ for
$i=1,\cdots,n-1$, and $\widetilde{y}=(y_1,\cdots,y_{n-1})^T$. Then
we obtain
$$
\dot{y}_i=\sum_{j=1}^{n-1}a_{ij}\big(f_j(y_j)-f_i(y_i)\big)-b_if_i(y_i).
$$

Since the subgraph $\mathcal{G}(\widetilde{\mathcal{A}})$ induced by
$\{1,\cdots,n-1\}$ is strongly connected, by Lemma 1, there exists
$\widetilde{\omega}=(\omega_1,\cdots,\omega_{n-1})^T$ such that
$\widetilde{\omega}^TL(\widetilde{\mathcal{A}})=0$. Define a
Lyapunov function
$\widetilde{V}(t)=\sum_{i=1}^{n-1}\omega_i\int_0^{y_i}f_i(s)\mathrm{d}s$.
Then
$$
\frac{\mathrm{d}\widetilde{V}(t)}{\mathrm{d}t}=-f(\widetilde{y})^T\mathrm{diag}(\widetilde{\omega})\big(L(\widetilde{\mathcal{A}})+\mathrm{diag}(\widetilde{b})\big)f(\widetilde{y})
=-f(\widetilde{y})^T\widetilde{B}f(\widetilde{y}),
$$
where
$\widetilde{B}=\big(\big(\mathrm{diag}(\widetilde{\omega})L(\widetilde{\mathcal{A}})+L(\widetilde{\mathcal{A}})^T\mathrm{diag}(\widetilde{\omega})\big)\big/2\big)+\mathrm{diag}(\widetilde{\omega})\mathrm{diag}(\widetilde{b})$.
From Lemma 1 and 2, $\widetilde{B}$ is positive definite. Denote the
smallest eigenvalue of it as $\lambda_1(\widetilde{B})>0$. Suppose
$\widetilde{V}(t)\not=0$, by utilizing Rayleigh-Ritz Theorem, we
obtain
$$
\frac{\mathrm{d}\widetilde{V}(t)}{\mathrm{d}t}=-\frac{f(\widetilde{y})^T\widetilde{B}f(\widetilde{y})}{f(\widetilde{y})^Tf(\widetilde{y})}\cdot\frac{f(\widetilde{y})^Tf(\widetilde{y})}{\widetilde{V}(t)^{\alpha}}\cdot
\widetilde{V}(t)^{\alpha}\le-\lambda_1(\widetilde{B})C_2\beta\widetilde{V}(t)^{\alpha}.
$$
Thereby, arguing as in Step 1 we get that $\widetilde{V}(t)$ and
$\widetilde{y}(t)$ will reach zero in finite time
$\widetilde{t}^*=V(0)^{1-\alpha}\big/\lambda_1(\widetilde{B})C_2\beta(1-\alpha)$.
By Lemma 1,
$L(\widetilde{\mathcal{A}})+\mathrm{diag}(\widetilde{b})$ is
positive definite and thus non-degenerate. Note that
$\widetilde{y}=-\big(L(\widetilde{\mathcal{A}})+\mathrm{diag}(\widetilde{b})\big)z
$, and then $\widetilde{y}=0$ yields $z=0$. Consequently, we obtain
$x=x_n1$ and the system solves a finite-time consensus problem with
the group decision value $x_n$.

Step 3. Suppose that $\mathcal{G}(\mathcal{A})$ has a spanning tree.

This general case can be proved by induction exactly as in \cite{7}.
We sketch the proof here for completeness. We introduce another
directed graph, denoted by $\mathcal{G}^c(\mathcal{A})$, consisting
of all strongly connected components $u_1,\cdots,u_k$ of
$\mathcal{G}(\mathcal{A})$, such that
$(u_i,u_j)\in\mathcal{E}(\mathcal{G}^c)$ if and only if there exist
$i'\in\mathcal{V}(u_i)$ and $j'\in\mathcal{V}(u_j)$ satisfying
$(i',j')\in\mathcal{E}(\mathcal{G})$.

The dynamics of agents corresponding to the vertex set of the root
of $\mathcal{G}^c(\mathcal{A})$ is not affected by others and the
local interconnection topology among them is strongly connected.
Hence by Step 1, the states of them will reach consensus in a finite
time. Denote the finial state by $x_0$. The induction step can
proceed along every path from root to leaves in
$\mathcal{G}^c(\mathcal{A})$ by employing Step 2 repeatedly. Since
there is a finite number of agents, the system solves a finite-time
consensus problem with finial state $x_0$. $\Box$

\bigskip
\noindent{\Large\textbf{4. Examples}}
\smallskip

In this section, to illustrate our theoretical results derived in
the above section, we will provide two concrete examples. Both are
seen to solve finite-time consensus problems.

\smallskip
\noindent\textbf{Example 1.}\quad In protocol (\ref{2}) for
$i=1,\cdots,n$, take
\begin{equation}
f_i(z)=a_i\mathrm{sign}(z)|z|^{c_i}+b_iz,\quad
z\in\mathbb{R}\label{5}
\end{equation}
where $a_i>0$, $b_i\ge0$, $0<c_i<1$, and $\mathrm{sign}(\cdot)$ is
the sign function defined as
$$
\mathrm{sign}(z)=\left\{\begin{array}{ll} 1,&z>0\\0,&z=0\\-1,&z<0
\end{array}\right..
$$

The above protocol is a generalization of some protocols introduced
in \cite{8,7}. In the sequel, we will show that it meets our
criteria (A1) and (A2).

\smallskip
\noindent\textbf{Claim 1.}\itshape \quad Suppose the interaction
topology $\mathcal{G}(\mathcal{A})$ has a spanning tree, then the
system (\ref{1}) solves a finite-time consensus problem when
protocol (\ref{5}) is applied.\normalfont

\smallskip
\noindent\textbf{Proof}. In view of Theorem 1, we need to verify the
assumptions (A1) and (A2) for (\ref{5}).

It is easy to see Assumption (A1) is satisfied. To prove (A2), let
$c=\max_{1\le i\le n}c_i$, $\alpha=2c\big/(1+c)$ and
$$
\beta=\min_{1\le i\le
n}\bigg\{\frac{a_i^2\cdot\min\big\{\big(\|L(\mathcal{A})\|_{\infty}\|x(0)\|_{\infty}\big)^{2c_i-\frac{2c(1+c_i)}{1+c}},\big(\|L(\mathcal{A})\|_{\infty}\|x(0)\|_{\infty}\big)^{2c_i-\frac{4c}{1+c}}\big\}}{2\cdot\max
\big\{\big(\frac{a_i}{1+c_i}\big)^{\frac{2c}{1+c}},\big(\frac{b_i}2\big)^{\frac{2c}{1+c}}\big\}}\bigg\}.
$$
Note that
$\mathrm{d}|z|^{k+1}\big/\mathrm{d}t=(k+1)\mathrm{sign}(z)|z|^k$ for
$k>0$. Hence, we have
$$
\frac{f_i(z)^2}{\big(\int_0^zf_i(s)\mathrm{d}s\big)^{\alpha}}=\frac{\big(a_i\mathrm{sign}(z)|z|^{c_i}+b_iz\big)^2}{\big(\frac{a_i}{1+c_i}|z|^{1+c_i}+\frac{b_i}{2}|z|^2\big)^{\alpha}}\ge
\frac{a_i^2|z|^{2c_i}}{\big(\frac{a_i}{1+c_i}\big)^{\frac{2c}{1+c}}|z|^{\frac{2c(1+c_i)}{1+c}}+\big(\frac{b_i}2\big)^{\frac{2c}{1+c}}|z|^{\frac{4c}{1+c}}}\ge\beta,
$$
where the last inequality follows from the fact
$2c_i-\frac{4c}{1+c}\le2c_i-\frac{2c(1+c_i)}{1+c}\le0$. $\Box$

\smallskip
\noindent\textbf{Example 2.}\quad In protocol (\ref{2}) for
$i=1,\cdots,n$, take
\begin{equation}
f_i(z)=\begin{cases}\ -a_i\mathrm{sign}(z)|z|^{c_i}\ln
|z|,&0<|z|\le e^{-1}\\
\ a_i\mathrm{sign}(z)|z|^{c_i},&|z|>e^{-1}\\
\ 0,&z=0
\end{cases}\label{6}
\end{equation}
where $a_i>0$, $0<c_i<2\big/3$.

We will show that (\ref{6}) is also a finite-time consensus
protocol.

\smallskip
\noindent\textbf{Claim 2.}\itshape \quad Suppose the interaction
topology $\mathcal{G}(\mathcal{A})$ has a spanning tree, then the
system (\ref{1}) solves a finite-time consensus problem when
protocol (\ref{6}) is applied.\normalfont

\smallskip
\noindent\textbf{Proof}. By straightforward calculation, it is easy
to see that Assumption (A1) holds. Note that
$a_i\mathrm{sign}(z)|z|^{c_i}\le f_i(z)\le
a_i\mathrm{sign}(z)|z|^{c_i/2}$, when $|z|\le e^{-1}$. Let
$c=\max_{1\le i\le n}c_i$, $\alpha=4c\big/(2+c)$,
$$
\beta_1=\min_{1\le i\le
n}\frac{a_i^2\big(\|L(\mathcal{A})\|_{\infty}\|x(0)\|_{\infty}\big)^{2c_i-\frac{4c(1+c_i)}{2+c}}}{2\big(\frac{a_i}{1+c_i}\big)^{\frac{4c}{2+c}}},\
\beta_2=\min_{1\le i\le
n}\frac{a_i^2\big(\|L(\mathcal{A})\|_{\infty}\|x(0)\|_{\infty}\big)^{2c_i-\frac{2c(2+c_i)}{2+c}}}{2\big(\frac{2a_i}{2+c_i}\big)^{\frac{4c}{2+c}}}
$$
and $\beta=\min\{\beta_1,\beta_2\}$. We may obtain (\ref{3}) with a
similar reasoning as in Claim 1. $\Box$

\smallskip

\noindent\textbf{Remark 3.}\itshape \quad It is noteworthy that both
examples above are not Lipschitz continuous at some points. Since
solutions reach $span\{1\}$ in finite time, there is no uniqueness
of solutions in backwards time. Therefore, the Lipschitz condition
must be violated (e.g. \cite{18} pp.8). \normalfont

To illustrate, we show simulation results involving four agents
using protocols (\ref{5}) and (\ref{6}) respectively over directed
network topology $\mathcal{G}$ as shown in Fig. 1. Note that
$\mathcal{G}$ in this case has a spanning tree, implying that the
conditions of Claim 1 and Claim 2 are satisfied. For simplicity, we
assume that $a_{ij}=1$ if $(j,i)\in\mathcal{E}(\mathcal{G})$, and
$a_{ij}=0$ otherwise. Take initial value $x(0)=(2,-1,3,-2)^T$.
Consider the following two cases of (\ref{5}) and (\ref{6}) respectively:\\
(i)\quad For $i=1,\cdots,n$, take
$f_i(z)=\mathrm{sign}(z)|z|^{3/4}+z$,
$z\in\mathbb{R}$.\\
(ii)\quad For $i=1,\cdots,n$, take
$$
f_i(z)=\begin{cases}\ -\mathrm{sign}(z)|z|^{1/2}\ln
|z|,&0<|z|\le e^{-1}\\
\ \mathrm{sign}(z)|z|^{1/2},&|z|>e^{-1}\\
\ 0,&z=0
\end{cases}.
$$
The simulation results are shown in Fig. 2 and Fig. 3, respectively.

\bigskip
\noindent{\Large\textbf{5. Conclusion}}
\smallskip

Finite-time consensus problems for continuous nonlinear multi-agent
systems are investigated in this paper. We propose general
sufficient criteria, under which the system achieves a consensus,
provided that the underlying interaction topology has a directed
spanning tree. We introduce new finite-time distributed protocols as
examples for using the criteria. Simulation results are given to
demonstrate the effectiveness of our theoretical results. Since we
only study the case when interconnection topologies are fixed, how
to consider the switching topology is our future research.

\begin{center}
\textbf{Figure captions}
\end{center}

Fig. 1\quad Directed network $\mathcal{G}$ of four vertices.
$\mathcal{G}$ has $0-1$ weights.

Fig. 2\quad Evolution of states over $\mathcal{G}$ with protocol
(\ref{2}) and (i).

Fig. 3\quad Evolution of states over $\mathcal{G}$ with protocol
(\ref{2}) and (ii).

\bigskip

\begin{center}
\setlength{\unitlength}{1mm}
\begin{picture}(60,60)
\put(2,58){1}\put(2,2){4}\put(58,2){3}\put(58,58){2}
\put(2,56){\vector(0,-1){50}}\put(5,59){\vector(1,0){51}}\put(59,56){\vector(0,-1){50}}\put(56,5){\vector(-1,1){51}}
\end{picture}
\end{center}
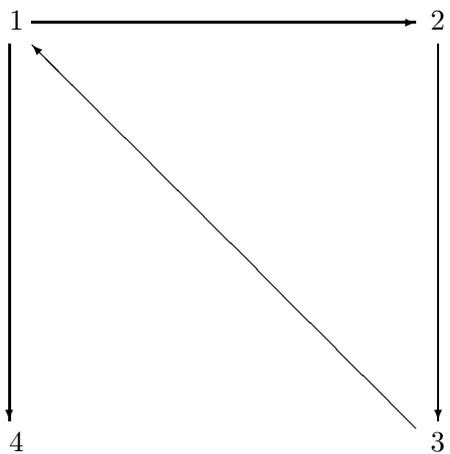
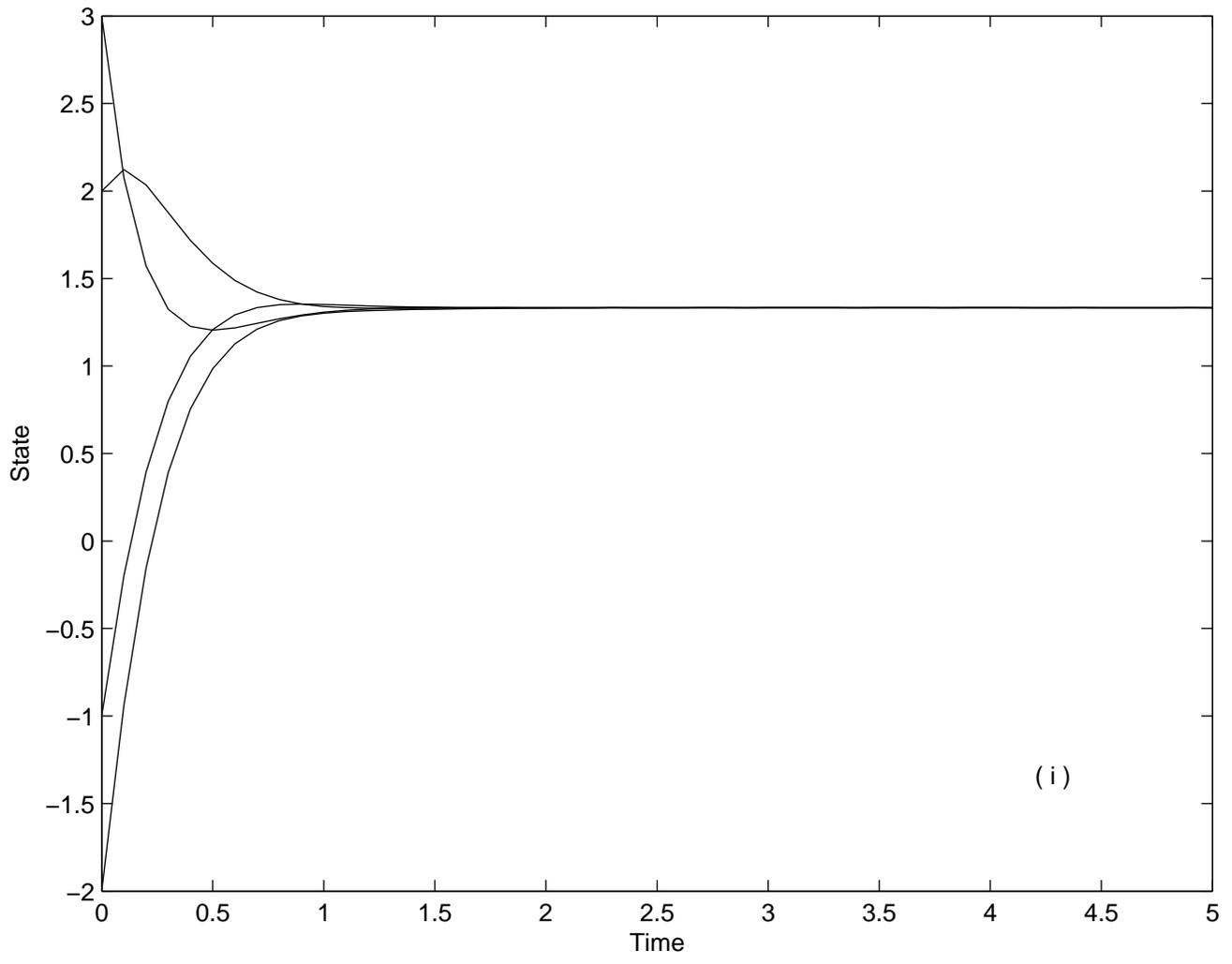
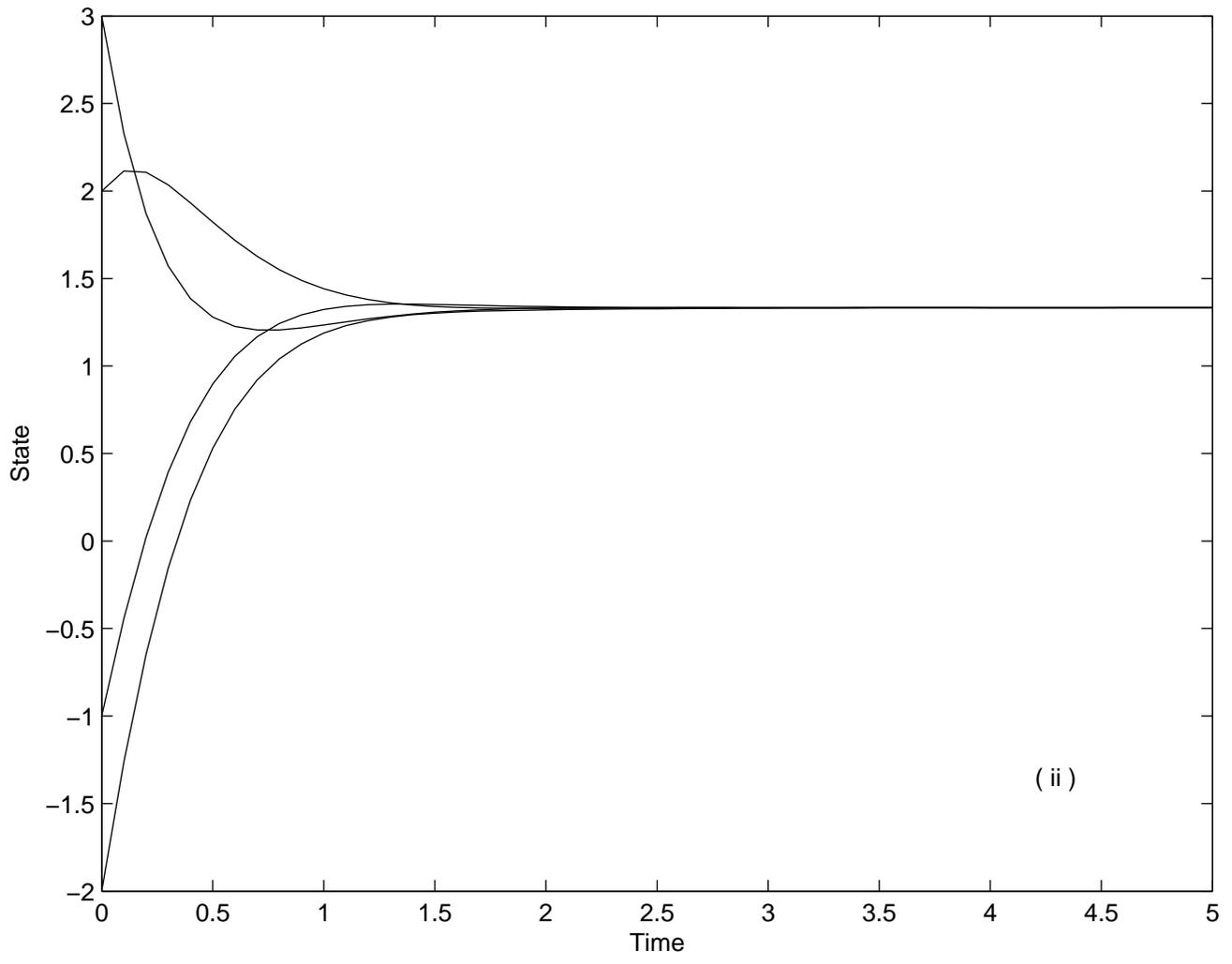
\end{document}